\documentclass[11pt,a4paper]{article}

\usepackage{jheppub}
\usepackage[utf8]{inputenc}
\usepackage{verbatim}
\usepackage{bbold}

\makeatletter
\def\@fpheader{Prepared for submission to JCAP}
\makeatother

\newcommand{\mn}{{\mu\nu}}

\newcommand{\be}{\begin{equation}}
\newcommand{\ee}{\end{equation}}
\newcommand{\om}{\Omega_m}
\newcommand{\omo}{\Omega_{m,0}}
\newcommand{\ol}{\Omega_\Lambda}
\newcommand{\ok}{\Omega_k}
\newcommand{\oko}{\Omega_{k,0}}
\newcommand{\omr}{\Omega_r}
\newcommand{\omro}{\Omega_{r,0}}
\newcommand{\gesim}{\,\raisebox{-0.4ex}{$\stackrel{>}{\scriptstyle\sim}$}\,}

\DeclareMathOperator{\Tr}{Tr}
\DeclareMathOperator{\Det}{Det}

\makeatletter
\DeclareRobustCommand{\rcite}[1]{%
  \rcite@aux#1,\@nil{#1}%
}
\def\rcite@aux#1,#2\@nil#3{%
  \if\relax#2\relax
    Ref.~\cite{#3}%
  \else
    Refs.~\cite{#3}%
  \fi
}
\makeatother
\title{Cosmological histories in bimetric gravity: A graphical approach}
\author[a,b]{E. M{\"o}rtsell}
\affiliation[a]{Oskar Klein Centre, Stockholm University,\\Albanova University Center\\ 106 91 Stockholm, Sweden}
\affiliation[b]{Department of Physics, Stockholm University\\
AlbaNova University Center\\ 106 91 Stockholm, Sweden}
\emailAdd{edvard@fysik.su.se}
\abstract{
The bimetric generalization of general relativity has been proven to be able to give an accelerated background expansion consistent with observations. Apart from the energy densities coupling to one or both of the metrics, the expansion will depend on the cosmological constant contribution to each of them, as well as the three parameters describing the interaction between the two metrics. Even for fixed values of these parameters can several possible solutions, so called branches, exist. Different branches can give similar background expansion histories for the observable metric, but may have different properties regarding, for example, the existence of ghosts and the rate of structure growth. In this paper, we outline a method to find viable solution branches for arbitrary parameter values. We show how possible expansion histories in bimetric gravity can be inferred qualitatively, by picturing the ratio of the scale factors of the two metrics as the spatial coordinate of a particle rolling along a frictionless track. A particularly interesting example discussed is a specific set of parameter values, where a cosmological dark matter background is mimicked without introducing ghost modes into the theory.
}

\begin{document}
\maketitle

\section{Introduction}\label{sec:intro}
We have recently celebrated the centennial of the advent of general relativity (GR) \cite{Einstein:1915ca}.  Remarkably enough, the relativistic field equations of GR still constitute state-of-the art in our understanding of gravitational phenomena. However, on the largest observed scales, in order to reach agreement with observations, we need to add additional components to the energy budget of the Universe: Dark matter, an invisible pressure less matter component needed to form and sustain galaxies and clusters of galaxies. Dark (or vacuum) energy, a nearly homogeneous component with negative pressure, needed to accelerate the current background expansion of the Universe. As long as these dark sectors have only been probed gravitationally, there still exists the possibility that a generalization of GR could be part of the explanation of the observations.

GR has been shown to be the unique second-order, local gravitational field equations derivable from an action built from a four-dimensional metric tensor \cite{Lovelock:1971yv}, and any generalization or modification of GR thus necessarily involves a breaking of one or more of these assumptions. Early attempts to modify GR included introducing a mass to the theory, effectively giving a mass to the particle mediating the gravitational force, the graviton \cite{Fierz:1939ix,Isham:1971gm}. However, it was long believed that massive gravity theories necessarily contained fatal ghost modes \cite{Boulware:1973my}. It was recently suggested that the inclusion of a second metric and a carefully constructed interaction between the two metrics of the theory could remove the ghost problem \cite{deRham:2010ik,deRham:2010kj}. That this is possible was proven in references~\cite{Hassan:2011hr,Hassan:2011tf,Hassan:2012qv}. Later work allowed for dynamics of also the second metric, so called bimetric gravity \cite{Hassan:2011zd}. 

In a quantum field theory picture, we can alternatively regard GR as a theory of a massless spin-$2$ field. Bimetric gravity then represents a natural extension of possible field theories to include also massive spin-$2$ interactions, i.e. the force of gravity is mediated by both massless and massive gravitons. 

When matter couples to both metrics, the theory can be made symmetric with respect to an interchange of the metrics \cite{Akrami:2013ffa,Akrami:2014lja,deRham:2014naa,deRham:2014fha,Enander:2014xga,Schmidt-May:2014xla,Comelli:2015pua,Gumrukcuoglu:2015nua,Solomon:2014iwa}. Here, we will constrain ourselves to the case where matter only couples directly to one of the metrics, for which the theory has interesting phenomenological consequences. The background cosmology of bimetric gravity includes an accelerated expansion without any explicit contribution from vacuum energy \cite{vonStrauss:2011mq, Volkov:2011an, Comelli:2011zm, Akrami:2012vf, Akrami:2013pna, Nersisyan:2015oha}. The evolution of small perturbations on top of a homogeneous background deviate from the growth of structure in GR. Specifically, in the linear approximation, scalar perturbations have been shown to grow increasingly fast on small scales \cite{Comelli:2012db,Khosravi:2012rk,Berg:2012kn,Fasiello:2013woa,Konnig:2013gxa,
Konnig:2014dna,Comelli:2014bqa,DeFelice:2014nja,Solomon:2014dua,Konnig:2014xva,Lagos:2014lca,
Enander:2015vja,Konnig:2015lfa,Aoki:2015xqa}. Similar instabilities have also been studied for tensor perturbations  \cite{Cusin:2014psa,Johnson:2015tfa,Amendola:2015tua,Fasiello:2015csa}. 
One way to remedy potential problems with these anomalies is to regard bimetric theory as an effective theory valid at low energies, i.e. at late cosmological epochs. By taking a GR limit of the theory, one can move the instabilities to very early epochs where a, currently unknown, more fundamental theory without the problematic instabilities is assumed to describe the growth of perturbations \cite{Akrami:2015qga}. 
However, since the instabilities only appear in the linear approximation, it has been argued that including non-linear effects will alleviate the effects of the instabilities \cite{Mortsell:2015exa}. This mechanism of restoring GR close to gravitational sources through non-linear terms, the Vainshtein mechanism, was proposed already in the early seventies \cite{Vainshtein:1972sx}. The Vainshtein mechanism is important also when constraining bimetric gravity using gravitational lensing observations \cite{Sjors:2011iv,Enander:2013kza}.

Even in the simplest bimetric generalization of GR, where matter only couples to one of the metrics, we still have four additional free parameters of the theory as compared to GR, allowing for a rich phenomenology of bimetric theory. Also, even for fixed values of the parameters of the theory, there are normally several different branches of solutions with different properties. This makes it demanding to assess the properties of such a large range of solutions, in a straightforward and methodical manner. The purpose of this paper is to remedy this by proposing a graphical method in which each set of parameter values of the bimetric theory generates a number of functions. When plotted, these functions directly give qualitative insight into the background expansion, the appearance of linear scalar instabilities, the Higuchi ghost and possible metric singularities, for all possible solution branches of the model.  

\section{Background equations}
The bimetric gravity Lagrangian $\mathcal{L}$ is given by
\begin{align}
\label{eq:HRaction}
\mathcal{L}=&-\frac{M_{g}^{2}}{2}\sqrt{-\det g}R_{g}-\frac{M_{f}^{2}}{2}\sqrt{-\det f}R_{f}\nonumber \\
&+m^{4}\sqrt{-\det g}\sum_{n=0}^{4}\beta_{n}e_{n}\left(\sqrt{g^{-1}f}\right)+\sqrt{-\det g}\mathcal{L}_\mathrm{m},
\end{align}
where $\mathcal{L}_m$ is the matter Lagrangian and $e_n$ are elementary symmetric polynomials given in, e.g., reference~\cite{Hassan:2011vm}. Varying the Lagrangian $\mathcal{L}$ with respect to the metrics $g_\mn$ and $f_\mn$, respectively, yields the equations of motion
\begin{align}
G^g_{\mu\nu}+m^{2}\sum_{n=0}^{3}\left(-1\right)^{n}\beta_{n}g_{\mu\lambda}Y_{\left(n\right)\nu}^{\lambda}\left(\sqrt{g^{-1}f}\right)&=\frac{1}{M_g^2}T_{\mu\nu},\\
G^f_{\mu\nu}+m^{2}\sum_{n=0}^{3}\left(-1\right)^{n}\beta_{4-n}f_{\mu\lambda}Y_{\left(n\right)\nu}^{\lambda}\left(\sqrt{f^{-1}g}\right)&=0.
\end{align}
Here, we have put $M_f=M_g$ through a rescaling of $f_\mn$ and the $\beta_n$ (see, e.g., references~\cite{Hassan:2011vm, Akrami:2015qga}). Assuming isotropy and homogeneity, our ans{\"a}tze for the metric fields $g_\mn$ and $f_\mn$ are 
\begin{align}
ds_{g}^{2}=-N_ g^2(t)dt^{2}+a_g^{2}(t)\left[\frac{R_0^2dr^2}{R_0^2-k r^2}+r^2d\Omega^2\right],\\
ds_{f}^{2}=-N_f^{2}(t)dt^{2}+a_f^{2}(t)\left[\frac{R_0^2dr^2}{R_0^2-k r^2}+r^2d\Omega^2\right].
\end{align}
Note that the metrics need to have the same radius of curvature, $\sqrt{k}/R_0$, in order to satisfy the equations of motion, and that we are free to put $N_ g(t)=1$ through a redefinition of the time coordinate $t$.
We define the ratio of the scale factors of the metrics, $r\equiv a_f/a_g$ and $H/H_0=\dot a_g/a_g$. Here, a dot denotes differentiation with respect to $\tau\equiv H_0 t$, and $H_0$ is the value of the Hubble parameter at the current epoch. We follow the approach in, e.g., \cite{vonStrauss:2011mq} and choose the so called dynamical branch of the Bianchi constraint. The resulting equations of motion can be written 
\be
\left(\frac{H}{H_0}\right)^2=\Omega+\ok+\omr,\label{eq:F1}
\ee
where
\be
\Omega\equiv\frac{8\pi G\rho}{3H_{0}^{2}},\label{eq:Omega}
\ee
\be
\ok\equiv-\frac{k}{H_{0}^{2}R_{0}^{2}a_g^{2}},\label{eq:Ok}
\ee
and 
\be
\omr\equiv\frac{B_{0}}{3}+B_{1}r+B_{2}r^{2}+\frac{B_{3}}{3}r^{3}.
\ee
Equation~(\ref{eq:F1}) is the $00$-component of the equations of motion for the $g$-metric.
Here,
\be
B_i\equiv \frac{m^2\beta_i}{H_0^2},
\ee
with $\beta_i$ being the cosmological constants and interaction parameters of the model. $\Omega$ is the energy density in units of the critical density, $\ok$ the effective curvature density and $\omr$ an effective interaction energy density, also containing the cosmological constant for the $g_{\mu\nu}$-metric, $B_0/3$.   
Alternatively, for $r\neq 0$, the $00$-component of the equations of motion for the $f$-metric can be written
\be
\left(\frac{H}{H_0}\right)^2=\ok+\frac{B_{4}}{3}r^{2}+B_{3}r+B_{2}+\frac{B_{1}}{3r}.\label{eq:F2}
\ee
The case of $r=0$ corresponds to the GR solution, with $B_0/3$ acting as a cosmological constant.
Equating equations (\ref{eq:F1}) and (\ref{eq:F2}) gives (again assuming $r\neq 0$)
\begin{equation}
\Omega (r)=-\frac{B_{3}}{3}r^{3}-\left(B_{2}-\frac{B_{4}}{3}\right)r^{2}-\left(B_{1}-B_{3}\right)r-\left(\frac{B_{0}}{3}-B_{2}\right)+\frac{B_{1}}{3r}.\label{eq:quartic}
\end{equation}
Note that we can use any pair of equations (\ref{eq:F1}), (\ref{eq:F2}),
and (\ref{eq:quartic}), but that not all three are independent. Also, the equations of motion are left unchanged
when simultaneously switching signs of $B_{1}$ and $r$ as well as $B_{3}$ and $r$.
For given values of the $B_i$ and the current curvature density $\oko$, we solve for the current energy density $\Omega_0$ and $r(t_0)=r_{0}$ by setting $H/H_0=1$ in equations~(\ref{eq:F1}) and (\ref{eq:F2})\footnote{Equation (\ref{eq:r0}) does not always have real solutions for
$r_{0}.$ For such parameter values, the only possible solution is the $r_0=0$ GR equivalent.}
\begin{eqnarray}
\frac{B_{3}}{3}r_{0}^{3}+B_{2}r_{0}^{2}+B_{1}r_{0}+\frac{B_{0}}{3}+\Omega_{0}+\Omega_{k,0}-1= 0,\label{eq:om0}\\
\frac{B_{4}}{3}r_{0}^{3}+B_{3}r_{0}^{2}+(B_{2}+\Omega_{k,0}-1)r_{0}+\frac{B_{1}}{3} = 0.\label{eq:r0}
\end{eqnarray}
Note that prior constraints on $\Omega_0$ and $\oko$ may differ from the GR case, since $\omr$ may contain terms with similar behaviour as, e.g., pressure less matter.  An explicit example is when we have $B_{1}=B_{3}=0$ (see section~\ref{sec:b1=b3=0}), in which case part of $\omr$ evolves as $\Omega$, and part as a cosmological constant. 

For given values of the parameters $[B_i, \Omega_0, \oko]$, in order to calculate the possible expansion histories, we first solve equation~(\ref{eq:quartic}) to obtain $r=r(a)$, which then in turn is used in equation~(\ref{eq:F1}) or (\ref{eq:F2}) to get $H=H(a)$. Though in principle straightforward,  equation~(\ref{eq:quartic}) has up to four different solutions (in our language corresponding to different {\em branches} of solutions), that in general has to obtained numerically. This makes it difficult to obtain a quick, qualitative understanding of the possible solutions. In the next section, we show how such an understanding can be accomplished using a graphical approach to the problem.

\section{Graphical approach}\label{sec:graphical}
\begin{figure}
\hspace{-2.5cm}\centerline{\includegraphics[width=10cm]{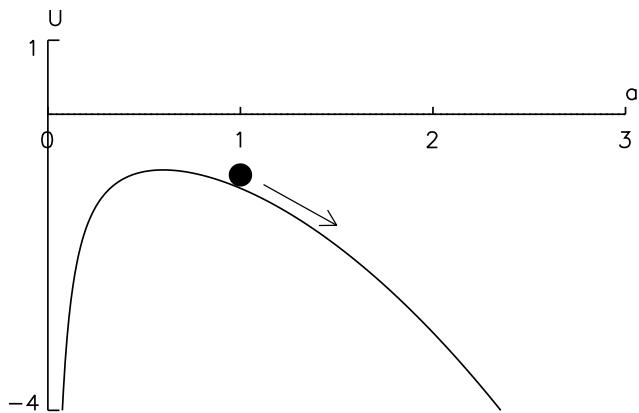}\hspace{-1.5cm}}
\caption{\label{fig:M=03L=07}The concordance cosmological model with $\om= 0.3$, $\ol= 0.7$ and zero spatial curvature. Our expansion history is represented by a particle rolling in from the left. Since $a_0=1$, we are currently living in a period of accelerated expansion.}
\end{figure}

In GR, it is possible to write the Friedmann equation in terms of the energies of a particle with total energy, $E$, rolling along a track, with the scale factor $a$ acting as coordinate. The kinetic energy is given by $K={\dot a}^2$ and the potential energy function can be written in terms of the scale factor only, $U=U(a)$ \cite{Mortsell:2016too}. 
In the standard model of the universe, at late epochs being dominated by pressure less matter, $\om$ and a cosmological constant, $\ol$ the potential and total energy and is given by \cite{Mortsell:2016too}
\begin{eqnarray}   
U =-\left(\frac{\om}{a}+\ol a^{2}\right),\\
E = \ok = 1-\om-\ol,
\end{eqnarray}
and $E=K+U$.
Plotting the potential energy function, we can immediately understand the allowed expansion histories of a given model, being given by the motion of a rolling particle that can maximally reach the height $\ok$.
The potential energy function of the so called concordance model, see e.g. \cite{Ade:2015xua}, with $\om= 0.3$, $\ol= 0.7$ and zero spatial curvature is shown in figure~\ref{fig:M=03L=07}. Note that the total energy $E=0$. The expansion history can now be depicted as a particle rolling in from the left, up the slope with decreasing velocity, corresponding to the decelerating matter dominated period. It rolls over the hill at $a\sim 0.5$, where the velocity is at its minimum value, and starts rolling down the slope with ever increasing velocity. This corresponds to the current accelerated phase when the cosmological constant is dominating the energy content of the Universe. 

For bimetric gravity, we are able to define corresponding quantities in terms of the ratio of scale factors $r$.  Using the chain rule, we write
\be
\dot r=\frac{dr}{d\tau}=\frac{dr}{d\Omega}\frac{d\Omega}{d\tau}=\frac{\dot\Omega}{\Omega'}.
\ee
Here, primes indicate differentiation with respect to $r$.
Combined with the energy conservation equation
\be
\dot\Omega = -3\left(\frac{H}{H_0}\right)\Omega (1+\omega),
 \ee 
 where $\omega$ is the total equation of state, we can write
 \be
 \dot r = -3\left(\frac{H}{H_0}\right)(1+\omega)\frac{\Omega}{\Omega'}.
 \ee
Defining the kinetic energy $K(r)\equiv\dot r^2$, and the potential energy
\be\label{eq:U}
U(r)\equiv -9\left(\frac{H}{H_0}\right)^2(1+\omega)^2\left(\frac{\Omega}{\Omega'}\right)^2,
\ee
the relation $E=K+U=0$ always holds. For the convenience of the reader, we write down the explicit form of the different factors of the potential energy function: 
 \begin{eqnarray}
\Omega(r)=-\frac{B_{3}}{3}r^{3}-\left(B_{2}-\frac{B_{4}}{3}\right)r^{2}-\left(B_{1}-B_{3}\right)r-\left(\frac{B_{0}}{3}-B_{2}\right)+\frac{B_{1}}{3r},\\
\Omega' (r)=-B_{3}r^{2}-2\left(B_{2}-\frac{B_{4}}{3}\right)r-\left(B_{1}-B_{3}\right)-\frac{B_{1}}{3r^2},\label{eq:omp}\\
\left(\frac{H}{H_0}\right)^2=\ok+\frac{B_{4}}{3}r^{2}+B_{3}r+B_{2}+\frac{B_{1}}{3r}.
\end{eqnarray}
Note that all quantities, except $\ok$, depend explicitly only on $r$. Since $a^{3(1+\omega)}=\Omega_0 /\Omega$, we can write
\be
\ok = \frac{\oko}{a^2}=\oko\left(\frac{\Omega}{\Omega_0}\right)^{2/3(1+\omega)}=\ok (r),
\ee
and the energy equation can be written in terms of $r$ only.
Notice that the only instance where the actual property of the energy density appears, through the equation of state $\omega$, is in the evolution of $\ok$ and as a multiplicative constant in the potential energy function. In the following, we will assume that the dominant energy component is pressure less matter, i.e., $\Omega=\om$ and $\omega=0$, although the method is applicable for arbitrary equations of state.

We can now plot the potential energy function~(\ref{eq:U}) to understand the possible evolution of $r$, and the corresponding evolution of $H(r)$, together with $\Omega (r)$, $\ok (r)$, $\omr (r)$ and $a(r)$. In addition, we can also plot quantities indicating the presence of scalar instabilities, ghosts and metric singularities. 
With a minimum of effort, we can thus quickly assess the validity of all possible branches of solutions.

\subsection{Scalar instabilities}
Using an ansatz where scalar perturbations are proportional to $e^{wt}$, we define the stability of linear perturbations as regions where $w$ is imaginary, i.e., where the equations of motion imply $w^2<0$. This guarantees that the perturbations are described by oscillating solutions, and not by exponentially increasing or decreasing functions.
To avoid scalar instabilities, we should thus have \cite{Konnig:2015lfa}
\begin{align}\label{eq:P}
P(r)&\equiv w^2\left(\frac{a}{k}\right)^2=\nonumber\\
&\frac{1}{H^2}\left[\frac{\Omega (1+\omega)}{\Omega' r^3}\left\{\frac{(r^2+1)(B_1-B_3 r)}{\Omega'}+\frac{r^2(B_1+4B_2 r+3B_3 r^2)}{B_1+2B_2 r + 3B_3 r^2}\right\}-1\right]< 0,
\end{align}
where we denote with $P(r)$ the scalar instability function. For given values of the $B_i$, negative values of $P(r)$ show that the model do not exhibit any linear scalar instabilities.
Note however that this is not a strict requirement on possible solutions, but rather an indication of when the linear approximation breaks down. As discussed in section~\ref{sec:intro}, 
non-linear effects may in fact alleviate the effects of the instabilities \cite{Mortsell:2015exa} through the Vainshtein mechanism \cite{Vainshtein:1972sx}. 

\subsection{Higuchi ghost}\label{sec:higuchi}
A {\em ghost} usually refers to a field with a negative kinetic energy. This renders the theory unstable, since the Hamiltonian is not bounded from below. As discussed in section~\ref{sec:intro}, the specific metric interaction potential in the bimetric Lagrangian (\ref{eq:HRaction}), is designed to guarantee the absence of the so called Boulware-Deser ghost, corresponding to a scalar propagating degree of freedom. However, it is still possible to have other ghosts in the theory. In reference \cite{Higuchi:1986py}, it was shown that the helicity-$0$ mode of a massive spin-$2$ field in a de Sitter background will be ghost like if $0<m^2<2H^2$. The corresponding bound to avoid the Higuchi ghost for bimetric gravity in general backgrounds was derived in reference~\cite{Fasiello:2013woa}, and corresponds to (see also \cite{Konnig:2015lfa})      
\be
3B_{3}r^{4}+2\left(3B_{2}-B_{4}\right)r^3+3\left(B_{1}-B_{3}\right)r^2+B_{1}\ge 0,
\ee
being equivalent to the condition $\Omega' \le 0$ [see equation~(\ref{eq:omp})].
When $r\rightarrow 0$, this enforces $B_1\ge 0$, or if $B_1=0$, $B_3\ge 0$. If also $B_3=0$, then $3B_2-B_0>0$. 
Plotting $\Omega (r)$, we can thus constrain our studies to parameter values for which the slope of $\Omega (r)$ is negative. 
Note that solutions fulfilling the Higuchi bound, exhibits a phantom behaviour of the interaction energy \cite{Konnig:2015lfa}. The implications of this for Big Rip scenarios are discussed in appendix~\ref{app:bigrip}. 

\subsection{Space time singularities}
We define the square root matrix $S$, appearing in the interaction terms of the bimetric Lagrangian (\ref{eq:HRaction}) as
\be
S\equiv\sqrt{g^{-1}f}.
\ee
In order to have non-singular metrics, scalars invariants derivable from this matrix, such as the trace and the determinant, should be finite. 
In reference~\cite{Deffayet:2011rh}, the curvature function $I$ is defined as 
\be
I\equiv \Tr{S^2} = f_\mn g^\mn =\frac{N_f^2}{N_g^2}+3\frac{a_f^2}{a_g^2}=N_f^2+3r^2.
\ee
The determinant of $S$ is given by
\be\label{eq:DetS}
\Det{S} =\frac{N_f}{N_g}\frac{a_f^3}{a_g^3}=N_fr^3.
\ee
The lapse of the $f$-metric is given by (assuming we choose the dynamical solution branch as opposed to the algebraic branch, see e.g. \cite{vonStrauss:2011mq})
\be
N_f = \frac{dr}{da}a+r=\frac{H_0}{H}\sqrt{-U}+r=3(1+\omega)\frac{\Omega}{\Omega'}+r.
\ee
In order to avoid space time singularities, we thus restrict ourselves to solution branches where $\Det{S}\neq 0$ and $\Det{S}\neq \pm\infty$.
 
\section{Examples}
In this section, we demonstrate the method outlined above with a few examples. Rather than trying to find solutions that provide good fits to the full set of observational data, the purpose is to show how the method can single out solutions that are eligible for a more careful analysis. 

The method involves plotting the potential energy function, $U(r)$, the Hubble rate $H^2(r)$, the matter density $\om (r)$, the interaction energy contribution $\omr (r)$ and possibly the scalar instability function $P(r)$ given in equation~(\ref{eq:P}) and the scalar invariant $\Det S$ of equation~(\ref{eq:DetS}). The evolution of $r$ corresponds to a particle rolling along the track given by the potential energy function, $U(r)$. Since the kinetic energy $K\geq 0$, and $E=K+U=0$, we are confined to regions where $U(r)\leq 0$. A particular branch of solution thus corresponds to a region where $U(r)\leq 0$, and the different branches are separated by regions where $U(r)\ge 0$. For a given solution branch to be a valid cosmological solution, we should have $\om\geq 0$ and $H^2\geq 0$ for the full range of $r$ traversed by the particle, and at some $r=r_0$ in this region, $H/H_0$ should be unity. The value $r=r_0$ correspond to our current epoch. As is the case for a rolling particle, the evolution is time symmetric in the sense that solutions can be represented by the particle rolling either to the right or to the left, that is, $r$ can either increase or decrease with time.

In order to avoid the Higuchi ghost, we need $\Omega' \leq 0$. Also, $\Det S$ should be finite in order to guarantee non-singular space times. 

If all of these requirements are fulfilled, we have a theoretically viable solution branch. To make a first rough assessment of the observational viability of the background expansion, we read off the current matter density and the interaction energy as the values of $\om$ and $\omr$ at the point where $H/H_0=1$. Usually, $\omo\sim 0.3$ and $\omro\sim 0.7$ indicates a fair agreement with observations but the fact that $\omr$ contains four different terms with different redshift evolution, mimicking different equations of state, allows also for other values to give an expansion history similar to the one observed. This possibility is realized, for example, in the $B_1=B_3=0$-model, see section~\ref{sec:b1=b3=0}.  

\subsection{$B_i=1$}
\begin{figure}
\hspace{-2.5cm}\includegraphics[width=10cm]{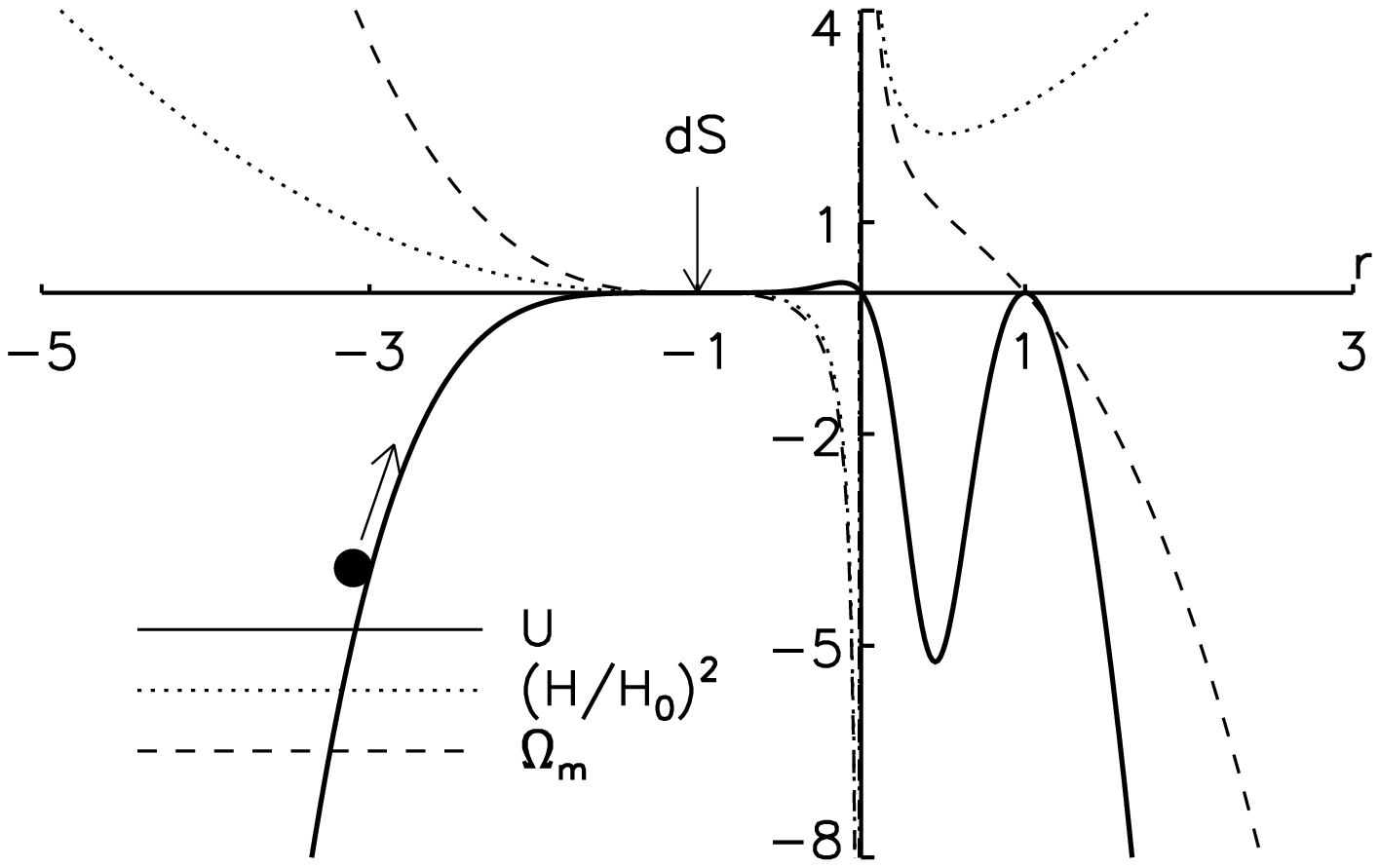}\hspace{-1.5cm}\includegraphics[width=10cm]{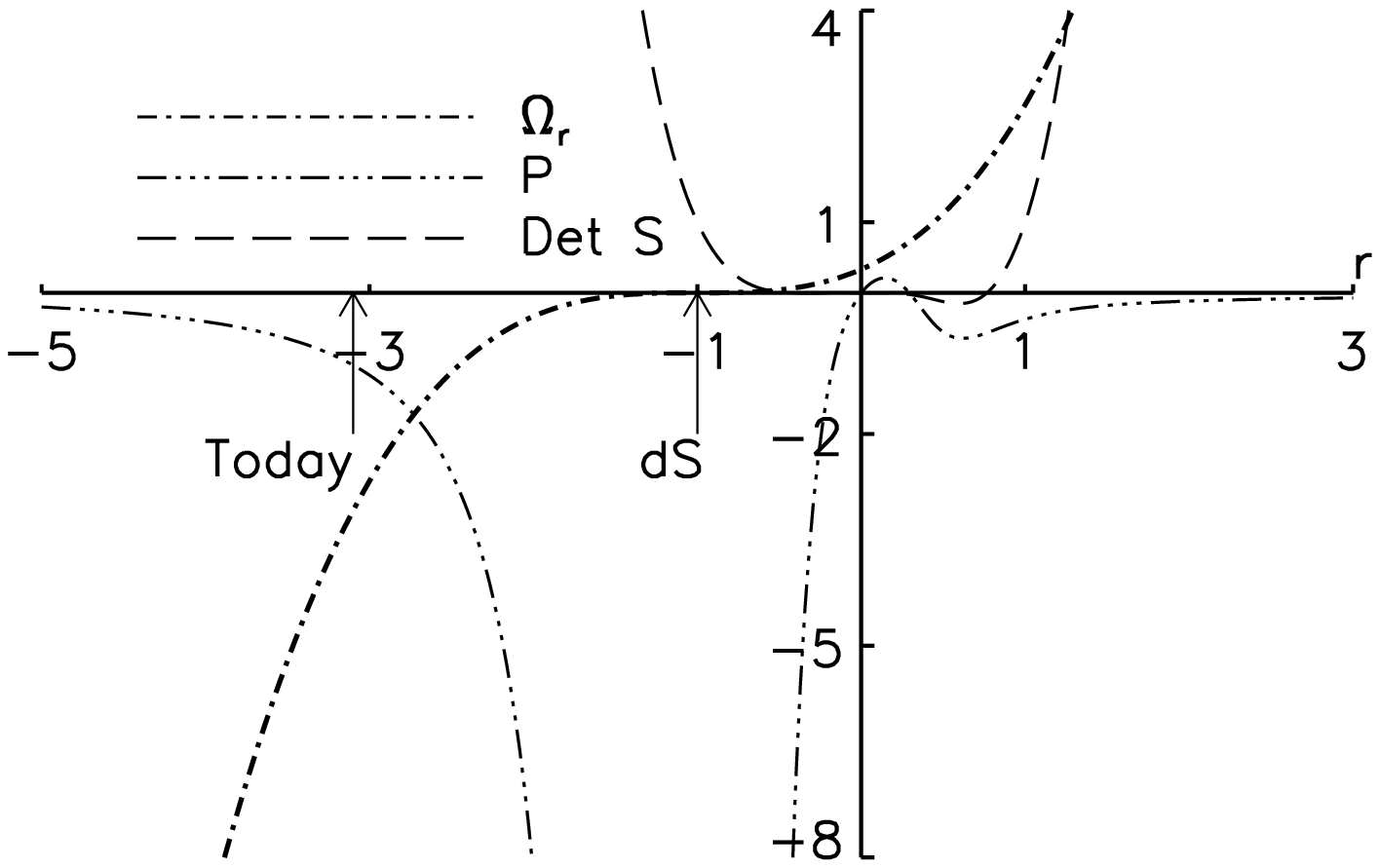}
\caption{\label{fig:bi=1} {\em Left:} The potential energy function, $U(r)$, Hubble rate $(H/H_0)^2(r)$ and matter density $\om (r)$ for the $B_i=1$-case. The only theoretically viable solution with $r\to -\infty$ in the infinite past and $r\to -1$ in the infinite future, is not expected to give a good fit to observational data since the matter density today (corresponding to $H/H_0=1$) is $\omo=4.1$. The value $r=-1$ correspond to the infinite future de Sitter (dS) phase. {\em Right:} The interaction energy contribution $\omr (r)$, scalar instability function $P(r)$ and $\Det S$ for the $B_i=1$-case. In the region $-\infty <r<-1$, $P(r)<0$, i.e., linear perturbations are stable and $\Det S$ is finite, i.e., there are no (background) metric singularities. The interaction energy is negative and approaches zero as $r\to -1$ in the infinite future.}
\end{figure}

Lacking knowledge of any possible prior probability of the $B_i$-parameters of the theory, our starting point is to put all $B_i$ to unity, and the spatial curvature to zero, $\oko=0$. In figure~\ref{fig:bi=1}, we show the graphical picture for this cosmology. Again, the evolution of $r$ is given by the motion of a particle, moving along a track given by the potential energy function $U(r)$. Since $E=0$, we are confined to regions of $r$ for which $U(r)<0$. 

From the left panel, na\"{\i}vely, it seems that we have two possibilities: $r\le -1$ or $0\le r\le 1$, both of them either with increasing $r$ with decreasing $\om$ or vice versa. However, we see that it is only the case of $r\le -1$ that can be normalized today to $(H/H_0)^2=1$, since for $0\le r\le 1$, we always have $(H/H_0)^2\gesim 2$. The right panel shows that $P(r)<0$, i.e., linear perturbations are stable and $\Det S$ is finite, i.e., there are no (background) metric singularities in the region $-\infty <r<-1$. Also, $\omr <0$, approaching zero as $r\to -1$. 
We can readily read off the fact that $(H/H_0)^2=1$ corresponds to $r_0\approx -3.1$ and a matter density today of $\om\approx 4.1$. Although theoretically viable, we thus do not expect the parameter set to supply a reasonable fit to observational data, because of the high value of $\om$. A possible caveat to this conclusion is that in principle, since $\oko=0$, the interaction terms acts as a negative energy density, and could cancel some of the effects of the large matter density on the background expansion.

\subsection{$B_1=B_4=1$}
\begin{figure}
\hspace{-2.5cm}\includegraphics[width=10cm]{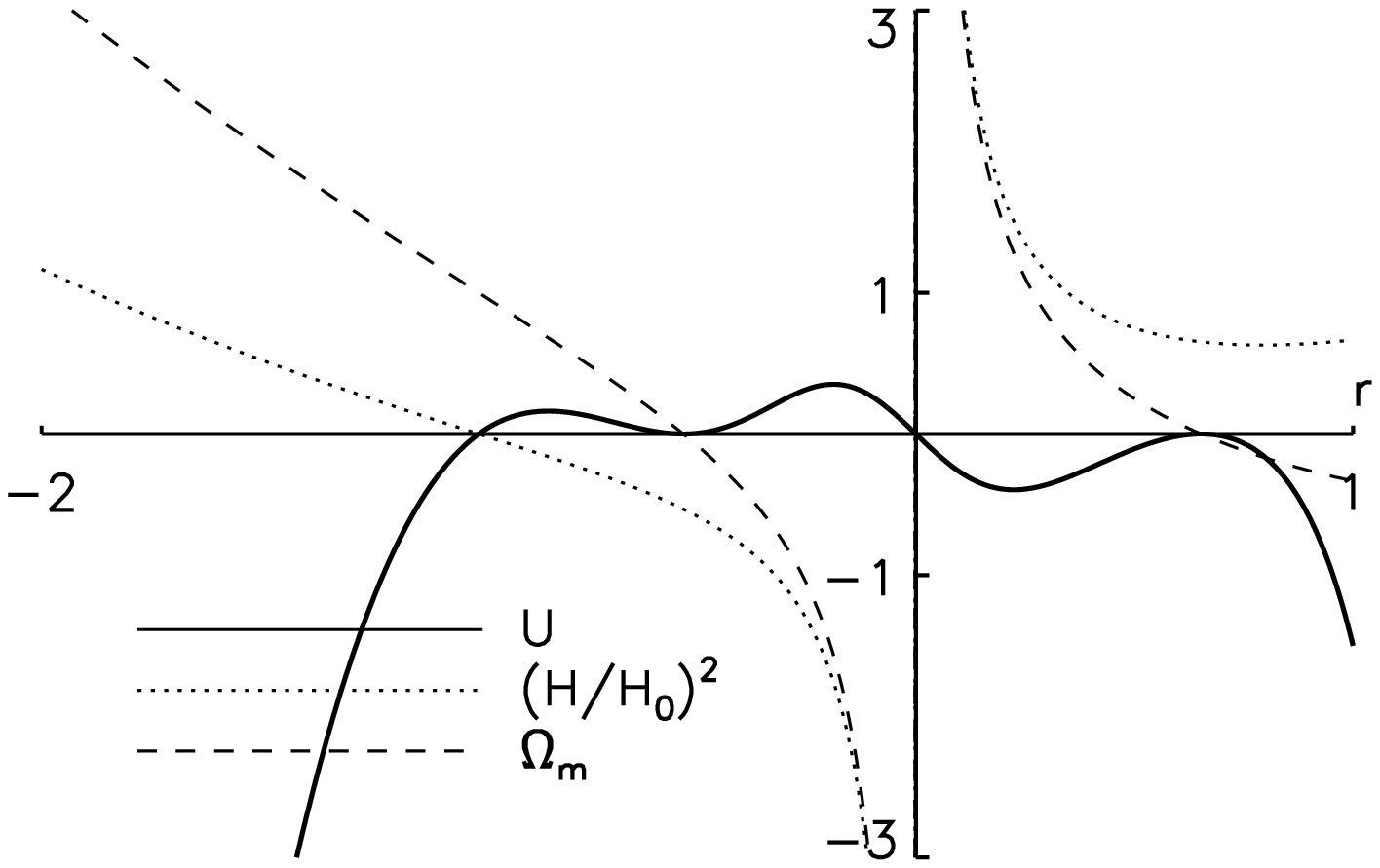}\hspace{-1.5cm}\includegraphics[width=10cm]{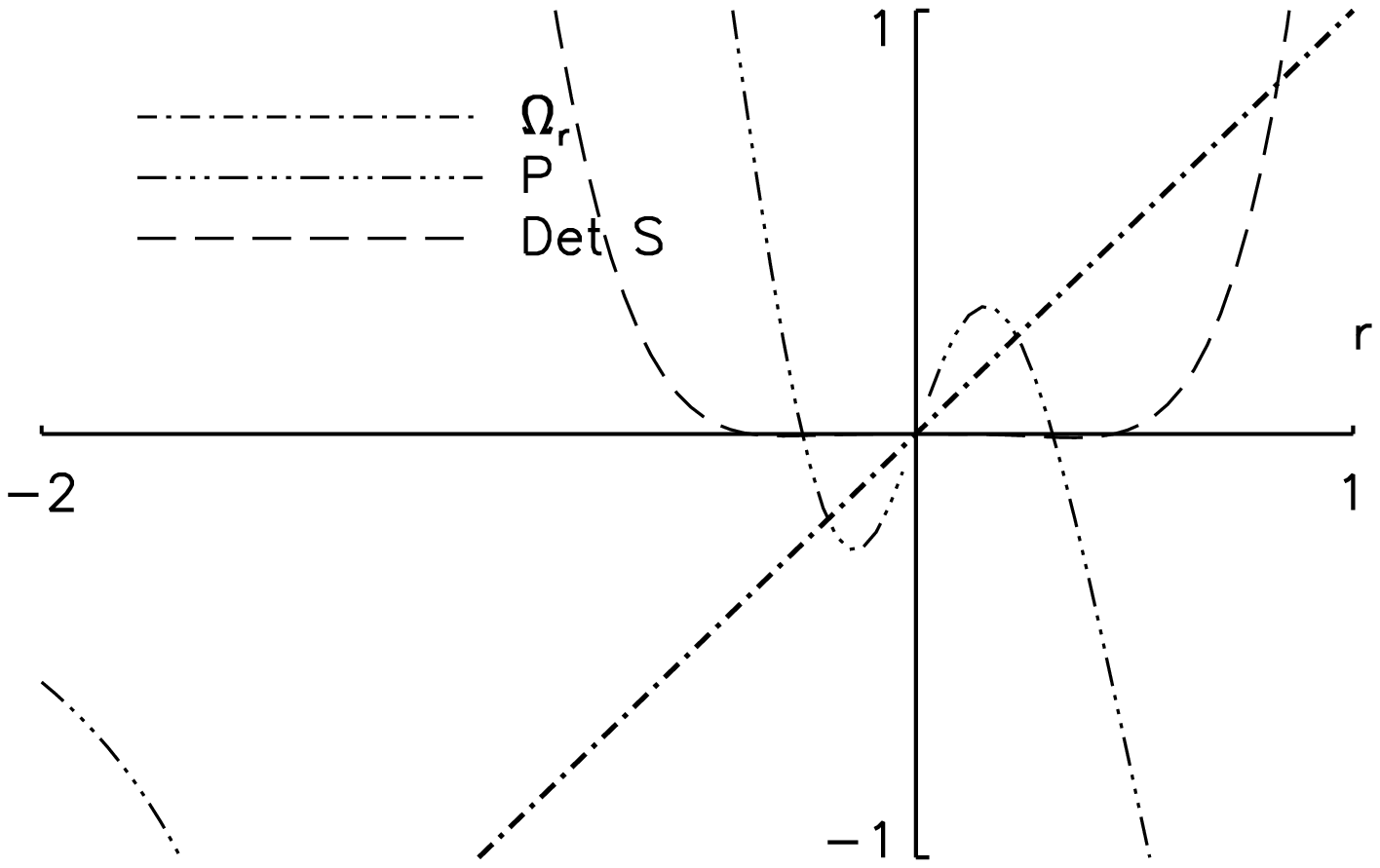}
\caption{\label{fig:b1b4} {\em Left:} The potential energy function, $U(r)$, Hubble rate $(H/H_0)^2(r)$ and matter density $\om (r)$ for the $B_1=B_4=1$-case. The solution branch with $0<r<\sim 0.65$ indicates a background evolution similar to our Universe. The branch with $-\infty<r<-1$ first expands and then contracts again.
{\em Right:} The interaction energy contribution $\omr (r)$, scalar instability function $P(r)$ and $\Det S$ for the $B_1=B_4=1$-case. In the region $-\infty <r<-1$, $P(r)<0$, i.e., linear perturbations are stable and $\Det S$ is finite, i.e., there are no (background) metric singularities. The interaction energy is negative and approaches zero as $r\to -1$ in the infinite future. The branch with $0<r<0.65$ displays a space time singularity at $r\approx 0.45$, as well as scalar instabilities for $r< 0.3$.}
\end{figure}

If we set $B_1=B_4=1$ and $\ok=B_0=B_2=B_3=0$, we obtain the graphical picture in the left panel of figure~\ref{fig:b1b4}. We have two possible branches of solutions: The first having $r\to 0$ in the infinite past, and $r\to \sim 0.65$ in the infinite future. The current matter density in this branch is $\omo=0.64$, too high by a factor of $\sim 2$, as compared to the standard cosmological model. 
The branch with $r\le -1$ shows an interesting behaviour. It represents a scenario where $r\to -\infty$, as $\om$ and $(H/H_0)^2\to\infty$, when going back in time. Going forward in time, $r$ increases as $\om$ and $(H/H_0)^2$ decreases. At $r= -1$, the universe comes to a halt and starts to contract again. This is the kind of solution that would be difficult to find without the graphical approach to the problem. 

The right panel of figure~\ref{fig:b1b4}, shows that the branch with $r\le -1$ does not have any scalar instabilities nor metric singularities, although the interaction energy $\omr$ is negative. In the branch with $0<r<0.65$, the interaction energy is increasing with time, i.e., it is displaying phantom properties. At early epochs ($r< 0.3$), scalar instabilities are present, invalidating linear perturbation theory. More importantly, the function $\Det S$ crosses zero at a finite $r$, showing that there is a space time singularity in the background space time, invalidating the solution branch. 

\subsection{$B_1=B_3=0$}\label{sec:b1=b3=0}
This model is of particular interest, since it was one of the first studied in terms of the background expansion of bimetric gravity \cite{vonStrauss:2011mq}. It has an effective rescaling of the matter density in the background equations, as well as an effective cosmological constant, and can thus provide an excellent fit to the observed expansion with baryonic matter only. Choosing $B_0=1.04$, $B_2=0.28$ and $B_4=1.00$, we obtain a current matter density of $\omo =0.05$ (in accordance with the estimated baryon density), but an effective background matter density of $0.3$, and a cosmological constant of $0.7$ as measured from the universal expansion history.

\begin{figure}
\hspace{-2.5cm}\includegraphics[width=10cm]{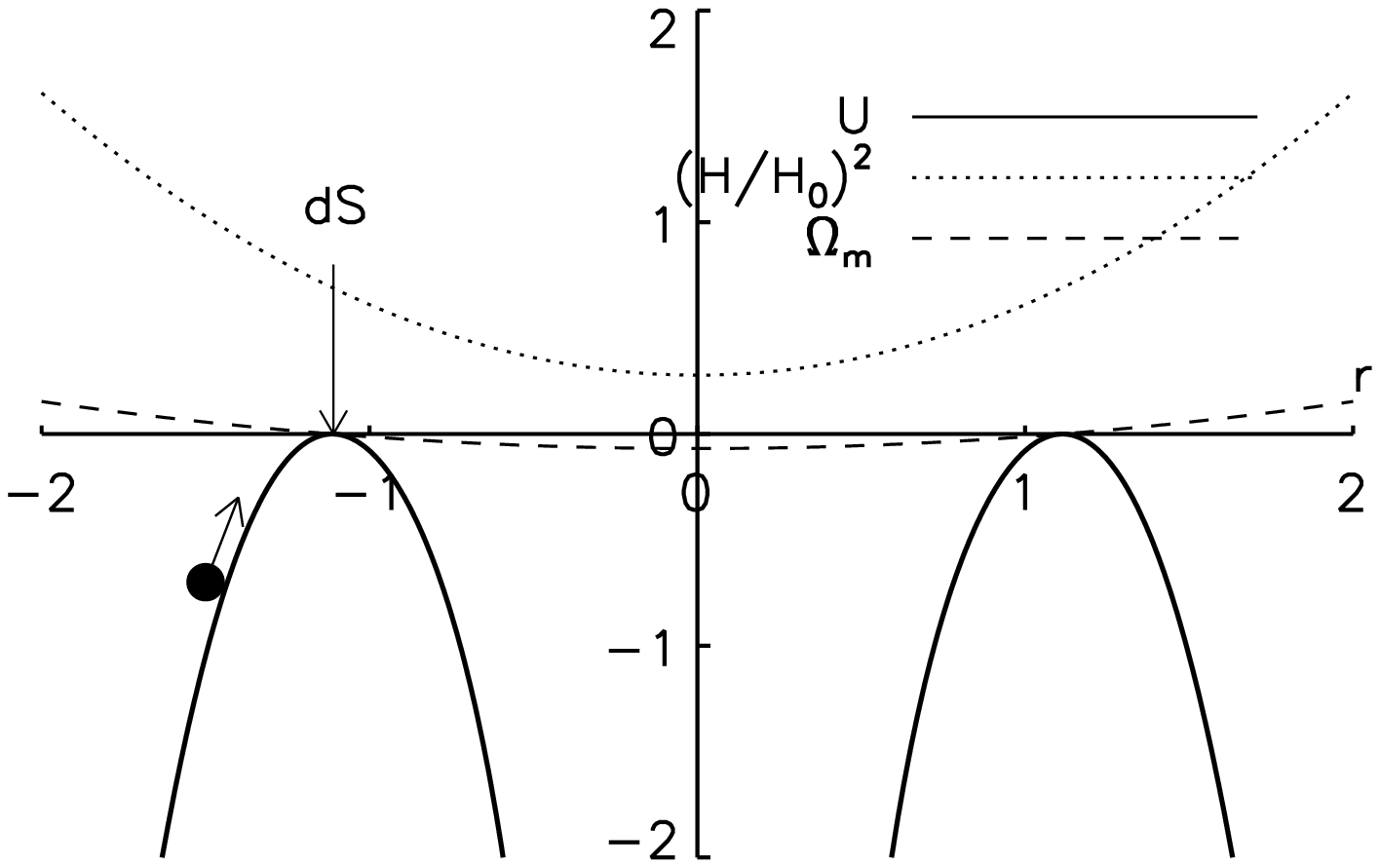}\hspace{-1.5cm}\includegraphics[width=10cm]{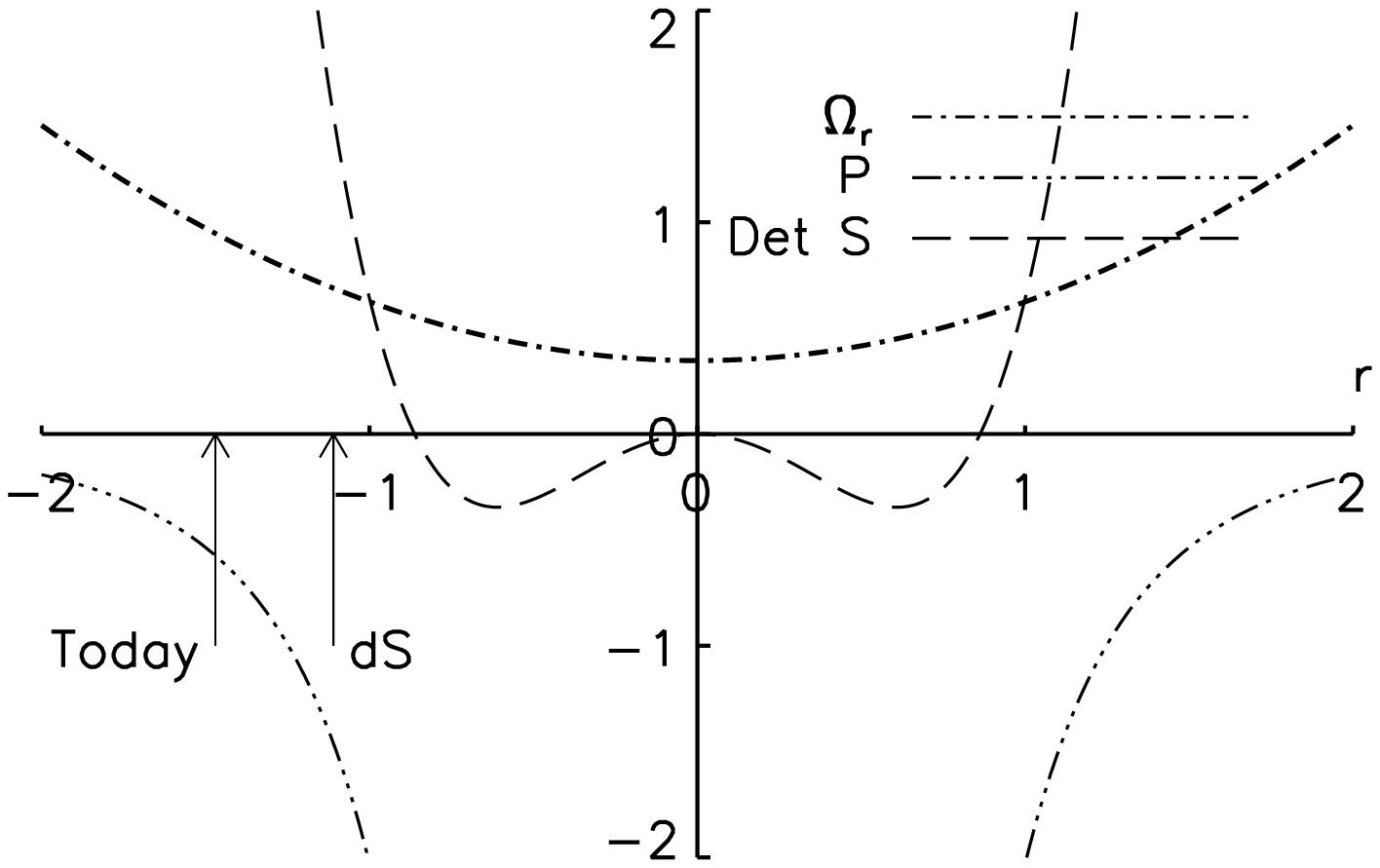}
\caption{\label{fig:b1b30} {\em Left:} The potential energy function, $U(r)$, Hubble rate $(H/H_0)^2(r)$ and matter density $\om (r)$ for the model with $B_1=B_3=0$, $B_0=1.04$, $B_2=0.28$ and $B_4=1.00$. Note the symmetry when changing the sign of $r$. Only the branch with $r\lesssim -1$ avoids the Higuchi ghost. Today, $r=-1.47$, and in the infinite future, $r=-1.11$, corresponding to a dS expansion phase.
{\em Right:} The interaction energy contribution $\omr (r)$, scalar instability function $P(r)$ and $\Det S$ for the same model. Linear perturbations are stable and $\Det S$ is finite for both solution branches.}
\end{figure}

Figure~\ref{fig:b1b30} shows the symmetry between the solutions with $r\lesssim -1$ and $r\gtrsim 1$, also evident from the fact that the analytical solution only involves quadratic powers of $r$ \cite{vonStrauss:2011mq}. Both branches have stable linear scalar perturbations and no background space time singularities. The difference however lies in the fact that the solution with negative $r$, corresponding to a particle rolling up the slope of $U(r)$ from the left, does not violate the Higuchi limit. This branch thus constitute a theoretically valid model, with a background expansion equivalent to one with a matter density $\omo=0.3$ and $\Omega_\Lambda=0.7$ but with baryonic matter $\Omega_b=0.05$ only. This particular solution was not considered in \cite{Konnig:2015lfa} since that analysis was restricted to positive $r$. The current value of $r$ is $r=-1.47$, whereas $r=-1.11$ corresponds to the value in the infinite future, where the model approaches a dS expansion phase.  

\section{Summary and discussion}
We have outlined a method to make a complete, qualitative evaluation of possible expansion histories, given the parameters of a bimetric cosmological model. The approach is to define a potential energy for a particle with zero total energy, where the ratio of scale factors of the two metrics, $r$, acts as spatial coordinate. 
The method also allows for evaluation of the existence of Higuchi ghosts, background space time singularities and linear scalar instabilities. 
Given the interaction parameters and cosmological constants, $B_i$, the equation of state $\omega$ of the energy content and the spatial curvature $\oko$, the validity of bimetric models can evaluated using the following recipe:
\begin{itemize}
\item Plot $U(r), H^2(r), \om (r)$ and $\omr (r)$ as described in section~\ref{sec:graphical}. 
\item Solutions are confined to regions where $U(r)$ is negative and finite and $[H^2,\om]\ge 0$. The current epoch corresponds to $H/H_0(r)=1$. 
\item The evolution of $r$ is given by the motion of a particle moving in the potential given by $U(r)$. To fulfill the Higuchi bound, only regions where $\Omega$ is decreasing with increasing $r$ are allowed.
\item Normally, observationally viable models include but are not limited to models having $\omo\sim 0.3$ and $\omro\sim 0.7$.
\item Points with $U=U'=0$ represent points in infinite future or past, see section~\ref{app:bigrip}. 
\end{itemize}
We applied the method to a small selection of parameter values, and found examples of re collapsing solutions ($B_i=1$), as well as solutions with the background expansion resembling the observed accelerated expansion, but plagued by space time singularities and scalar linear instabilities ($B_1=B_4=1$). Most interestingly, we revisited a model being able to mimic the background evolution of a $\omo=0.3$ and $\Omega_\Lambda=0.7$ model using baryonic matter only ($B_1=B_3=0$). Allowing for negative values of $r$, the model has stable linear scalar perturbations, no metric singularities at the background level and is not plagued by the Higuchi ghost. It would be of interest to study closer this, and related models, with respect to, e.g., the full evolution of scalar and tensor perturbations, the impact of the energy scaling on radiation dominated epochs and the Vainshtein mechanism.    

\acknowledgments
Support for this study from the Swedish Research Council is acknowledged. Also, I am grateful to Jonas Enander for his careful reading of the manuscript and valuable comments, and the anonymous referee for useful suggestions.
 
\bibliographystyle{JHEP}
\bibliography{bibliography}

\appendix
\section{Big Rip}\label{app:bigrip}
In general relativity, a phantom energy component ($w<-1$) will, if at some point dominating the total energy budget, make the scale factor of universe become infinitely large in a finite time. Such a scenario is usually referred to as a Big Rip and represents a space time singularity at the background level. As noted in section~\ref{sec:higuchi}, any expanding bimetric solution branch fulfilling the Higuchi bound will have its interaction energy growing with time, i.e., effectively having an effective equation of state $w<-1$. The question then arises if we always will have a future Big Rip in these solutions, or if the fact that the interaction energy approaches a constant value in the infinite future, i.e., that $w\to -1$, avoids the future space time singularity. We start by writing
\be
dt = \frac{dr}{\dot r}=-\frac{dr \Omega'}{3H(1+\omega)\Omega}=-\frac{d\Omega}{3H(1+\omega)\Omega},
\ee
where the sign has been chosen to have the density decreasing with increasing time. 
The question is whether integrating $\Omega$ from infinity to zero yields a finite or infinite time,
\be
t=-\int_0^\infty\frac{d\Omega}{3H(1+\omega)\Omega}.
\ee
Since $H\propto\sqrt{\Omega}$ as $\Omega\to\infty$, the integral stays finite in this limit. For $\Omega\to 0$, it is easy to see that $H^2=H_{\rm dS}^2(1+k\Omega )$ to first order in $\Omega$. This means that in the interval $\Omega=[0,\epsilon]$, we have
 \be
t\propto-\int_0^\epsilon\frac{d\Omega}{\Omega\sqrt{1+k\Omega}}=\left[2\,{\rm atanh}\,\sqrt{1+k\Omega}\right]_0^\epsilon=\infty,
\ee
i.e., it will take an infinite time to get to $\Omega =0$, corresponding to $a=\infty$, and we will not have a Big Rip.

As a side note; whether it takes finite or infinite time to reach $U(r)=0$, depends on the value of $U'$ at that point, as can be understood from the rolling particle picture. Using ${\dot r}^2=-U$, we can write 
\be
\frac{dr}{\sqrt{-U}}=dt,
\ee
and defining $\epsilon\equiv r_1-r$, the time it takes to go from $r$ to $r_1$ where $U(r_1)=0$ is
\be
t=\int_0^\epsilon\frac{d\epsilon}{\sqrt{-U}}.
\ee   
Taylor expaning around $r_1$ gives 
\be
U=-U'\epsilon+\frac{U''\epsilon^2}{2}+\mathcal{O}(\epsilon^2).
\ee 
For $U'\neq 0$, we get $t=\sqrt{\epsilon /U'}/2$ and for $U'=0$, $t=\infty$.

\end{document}